\documentclass[
11pt,
headings=normal,
numbers=noenddot
]{scrartcl}

\usepackage[a4paper,margin=1in]{geometry}

\usepackage{amsmath,amssymb}
\usepackage{graphicx}
\usepackage{booktabs}
\usepackage{hyperref}
\usepackage{microtype}
\usepackage{caption}
\usepackage{subcaption}

\hypersetup{
    colorlinks,
    linkcolor={red!50!black},
    citecolor={blue!50!black},
    urlcolor={blue!80!black}
}

\usepackage[bitstream-charter]{mathdesign}
\DeclareSymbolFont{usualmathcal}{OMS}{cmsy}{m}{n}
\DeclareSymbolFontAlphabet{\mathcal}{usualmathcal}

\usepackage{scrlayer-scrpage}
\ohead{\pagemark}

\usepackage[ruled,vlined,linesnumbered]{algorithm2e}
\DontPrintSemicolon
\SetKwInOut{KwIn}{Input}
\SetKwInOut{KwOut}{Output}
\SetKwComment{tcp}{// }{}
\SetAlFnt{\small}

\usepackage{graphicx}
\usepackage{tikz}
\usetikzlibrary{arrows.meta,positioning,fit,matrix,calc}

\usepackage{amsmath,amssymb}
\usepackage{booktabs}
\usepackage{listings}
\usepackage{xcolor}
\usepackage{physics}
\usepackage{float}
\usepackage{hyperref}
\usepackage{soul}
\usepackage{tikz}
\usetikzlibrary{arrows.meta, positioning, shapes.geometric, calc, fit}

\definecolor{scipostdeepblue}{HTML}{002B49}

\usepackage{tikz}
\usetikzlibrary{arrows.meta, positioning, shapes.geometric, calc, fit, matrix}

\tikzset{
  pnode/.style={
    draw,
    rounded corners=2pt,
    align=center,
    font=\small,
    text width=0.28\textwidth,
    minimum height=9mm,
    inner xsep=6pt,
    inner ysep=4pt
  },
  psmall/.style={
    draw,
    rounded corners=2pt,
    align=center,
    font=\small,
    text width=0.22\textwidth,
    minimum height=9mm,
    inner xsep=6pt,
    inner ysep=4pt
  },
  pwide/.style={
    draw,
    rounded corners=2pt,
    align=center,
    font=\small,
    text width=0.86\textwidth,
    minimum height=9mm,
    inner xsep=6pt,
    inner ysep=4pt
  },
  parrow/.style={-Latex, line width=0.6pt},
  pgroup/.style={draw, rounded corners=3pt, inner sep=6pt}
}

\begin{document}

\begin{center}{\Large \textbf{\color{scipostdeepblue}{
A geometry-first tutorial for time-resolved morphological analysis with \texttt{PyPETANA}
}}}\end{center}

\begin{center}\textbf{
Benjamin Evert Himberg\textsuperscript{1,2$\dagger$} and
Sanghita Sengupta\textsuperscript{3$\star$}
}\end{center}

\begin{center}
{\textbf 1} Department of Physics, University of Vermont, Burlington, VT, USA\\
{\textbf 2} Materials Science Program, University of Vermont, Burlington, VT, USA\\
{\textbf 3} Department of Chemistry, Brandeis University, Waltham, MA, USA\\[0.5em]
$\star$ \href{mailto:ssengupta@brandeis.edu}{\small ssengupta@brandeis.edu},\quad
$\dagger$ \href{mailto:bhimberg@uvm.edu}{\small bhimberg@uvm.edu}
\end{center}

\section*{\color{scipostdeepblue}{Abstract}}

We present a step-by-step, reproducible tutorial for \texttt{PyPETANA}, an open-source Python framework for geometry-first, time-resolved quantification of evolving morphology from image data. Starting from time-lapse video input, the tutorial demonstrates how to extract binary masks, compute time-resolved geometric observables including area, perimeter, circularity, and effective fractal dimensions, and analyze their temporal evolution. The workflow emphasizes direct reconstruction of morphology from images without assuming microscopic growth mechanisms. In addition to compactness-sensitive geometric descriptors, the framework supports multiscale boundary analysis through supersampled box-counting methods applied to filled morphologies and finite-width boundary bands. The benchmark suite further demonstrates applicability to invasive tumor morphologies and multiscale boundary evolution in time-resolved cancer-growth interfaces. This tutorial accompanies the computational workflow underlying \href{https://arxiv.org/abs/2602.05958}{arXiv:2602.05958} and provides a reproducible foundation for geometry-based analysis of evolving non-equilibrium morphologies.

\vspace{\baselineskip}

\vspace{10pt}
\noindent\rule{\textwidth}{1pt}
\tableofcontents
\noindent\rule{\textwidth}{1pt}
\vspace{10pt}

\section{Introduction}
\label{sec:intro}

Morphological pattern formation is ubiquitous in non-equilibrium systems, spanning bacterial colony growth, viscous fingering, dendritic solidification, electrodeposition, diffusion-limited aggregation, and active-matter aggregation \cite{benjacob1990nonequilibrium,langer1980instabilities,saffman1958penetration,witten1981diffusion,marchetti2013hydrodynamics,hallatschek2023proliferating,cross1993pattern}. Across these settings, morphology itself is often an information-rich observable: it reflects the integrated effects of transport, instabilities, and interactions, even when the underlying microscopic dynamics are only partially known or difficult to parameterize. In bacterial colonies in particular, classic studies established that growth on agar can produce distinct morphological classes, including compact, branching, and diffusion-limited forms, depending on nutrient availability and substrate conditions \cite{matsushita1990,matsushita1999}. Related morphology-driven approaches have also been used to characterize invasive tumor interfaces and multiscale roughening in cancer-growth morphologies\cite{brouwer2023,10.1371/journal.pone.0109784}. These observations have made colony morphology a useful probe of non-equilibrium growth and transport-limited pattern formation \cite{benjacob1997snowflake}.

Advances in microscopy, time-lapse imaging, and automated analysis have made large image sequences increasingly accessible, shifting the primary challenge from data acquisition to reproducible quantification. Recent workflows have enabled time-resolved measurements of colony growth, morphology, and biofilm structure from image data \cite{bar2020coltapp,bottura2022channels}, while machine-learning-based approaches are increasingly used for segmentation and classification tasks \cite{spahn2022deepbacs}. At the same time, a growing ecosystem of software tools addresses related problems in microbial image analysis. BiofilmQ~\cite{Hartmann2021} provides comprehensive 3D biofilm quantification, MicrobeJ~\cite{Ducret2016} is widely used for single-cell and colony shape analysis, ABC3D~\cite{Rooney2026} focuses on fractal and textural analysis of biofilm morphology, and Cellpose~\cite{Stringer2021} provides deep-learning-based segmentation increasingly adopted across the field.

A central challenge is therefore not simply the availability of image data, but the construction of a reproducible, time-resolved geometric description that can be compared consistently across experiments and analysis pipelines. Existing workflows often prioritize endpoint characterization, segmentation, or classification, while the mapping from pixel-level representations to reported observables may remain implicit or strongly dependent on preprocessing and threshold choices. As a result, it can be difficult to disentangle genuine morphological evolution from variations introduced by parameter tuning or analysis strategy.

\texttt{PyPETANA} is a geometry-first framework for extracting time-resolved morphological observables directly from image data. The workflow treats morphology as a dynamical geometric object reconstructed from segmented images while maintaining a strict separation between image preprocessing and the mathematical definition of observables. Rather than introducing a new microscopic growth model or segmentation algorithm, the primary contribution of \texttt{PyPETANA} lies in providing a reproducible measurement protocol for morphology that remains agnostic to microscopic growth mechanism.

The framework combines standard OpenCV-based image operations with a deterministic workflow architecture for batch analysis. In particular, \texttt{PyPETANA} implements: (i) a parameter-record interpolation scheme enabling temporally smooth execution from a sparse set of interactively selected key frames; (ii) a strict separation between GUI orchestration and the numerical backend, ensuring that interactive preview and batch execution call identical routines; and (iii) a reproducibility framework defined by the input media, exported \texttt{records.json} snapshots, and the Zenodo-tagged code version. Within this framework, all reported observables are deterministic and reproducible across re-runs and users.

Beyond standard global descriptors such as area and perimeter, the workflow emphasizes boundary-sensitive and multiscale observables, including circularity and effective fractal dimensions computed for both the filled morphology and a controlled boundary band. To reduce discretization and grid-alignment artifacts in box-counting analysis, \texttt{PyPETANA} implements a supersampled padding strategy that reports fit diagnostics alongside extracted scaling exponents.

This article provides a practical and reproducible guide to running \texttt{PyPETANA} on time-resolved image data. The primary outputs are time series of geometric observables including area $A(t)$, perimeter $P(t)$, circularity $C(t)$, and effective fractal dimensions for the boundary and filled morphology, $D_b(t)$ and $D_f(t)$, respectively. The physical interpretation of these observables in the context of bacterial colony growth is discussed separately in \href{https://arxiv.org/abs/2602.05958}{arXiv:2602.05958}. The present article instead focuses on the computational workflow, software architecture, and reproducible extraction of geometric observables from evolving morphologies.

A run is reproducible given: (i) the input media (video or ordered image frames), (ii) the exported GUI snapshot JSON (\texttt{records.json}), and (iii) the Zenodo-tagged code version. Additional lightweight run metadata are embedded in the header of \texttt{observables.dat} via a \texttt{meta\_json=\{...\}} entry. The following sections provide an overview of the codebase, installation procedure, and the full analysis workflow implemented in \texttt{PyPETANA}.



\section{Package overview and installation}
\label{sec:package}

\subsection{Package information}
\label{sec:package_info}

\texttt{PyPETANA} is an open-source Python codebase released under the
GNU General Public License v3.0 (GPL-3.0)~\cite{sanghita_pypetana_2024}.
The present article documents \texttt{PyPETANA} version~2.0.0, archived
on Zenodo, which represents a substantial architectural and functional
revision relative to earlier prototype versions of the framework.

\subsection*{Installation}

\paragraph{Source code acquisition.}
The \texttt{PyPETANA} source code may be obtained either by cloning the
development repository hosted on GitHub or by downloading the archived
source release associated with the cited Zenodo version.

Users wishing to clone the development repository may do so via:
\begin{lstlisting}[language=bash]
git clone git@github.com:sanghita211/pypetana.git
\end{lstlisting}

Alternatively, the archived source code may be downloaded directly from
the Zenodo record corresponding to \texttt{PyPETANA} version~2.0.0.

\subsection{Setup and execution}

\texttt{PyPETANA} uses Pixi as its package and environment manager to
provide a reproducible runtime environment across platforms.

\vspace{\baselineskip}

\noindent Install Pixi on Mac/Linux using:
\begin{lstlisting}[language=sh]
curl -fsSL https://pixi.sh/install.sh | sh
\end{lstlisting}

\vspace{\baselineskip}

\noindent Install Pixi on Windows using:
\begin{lstlisting}[language=sh]
powershell -ExecutionPolicy Bypass -c "irm -useb https://pixi.sh/install.ps1 | iex"
\end{lstlisting}

\vspace{\baselineskip}

\noindent After installation, open a new terminal or PowerShell session,
navigate to the \texttt{PyPETANA} source directory, and execute:
\begin{lstlisting}[language=sh]
pixi run pypetana
\end{lstlisting}

\vspace{\baselineskip}

Pixi will automatically construct the required runtime environment and
launch the \texttt{PyPETANA} graphical user interface (GUI). The GUI
serves as the orchestration layer for the full analysis pipeline. The
following section introduces the GUI workflow and its role in interactive
segmentation validation and batch execution.


\section{Graphical User Interface (GUI)}
\label{sec:interfaces}

\subsection{Launching the GUI}
\label{sec:launch_gui}

\texttt{PyPETANA} is distributed as a single application whose primary
user-facing interface is a graphical user interface (GUI). After activating
the reproducible runtime environment specified by \texttt{pixi.toml} and
\texttt{pyproject.toml}, the GUI may be launched via:

\begin{lstlisting}
cd pypetana
pixi shell
pypetana
\end{lstlisting}

The \texttt{pypetana} entry point invokes \lstinline{main()} in the
application module, which initializes the GUI and enters the application
event loop. The GUI provides the default interface for interactive parameter
selection, segmentation validation, and batch execution.

\subsection{GUI workflow}
\label{sec:gui_workflow}

The GUI provides structured access to the full analysis pipeline:
\begin{enumerate}
    \item load a time-lapse video or image directory;
    \item preview frames and validate preprocessing and segmentation using
    real-time overlays;
    \item store parameter snapshots (\texttt{records}) at selected frames;
    \item execute batch extraction using interpolated per-frame parameters,
    generating numerical outputs and optional diagnostic exports.
\end{enumerate}

\begin{figure}[H]
    \centering
    \includegraphics[width=0.9\linewidth]{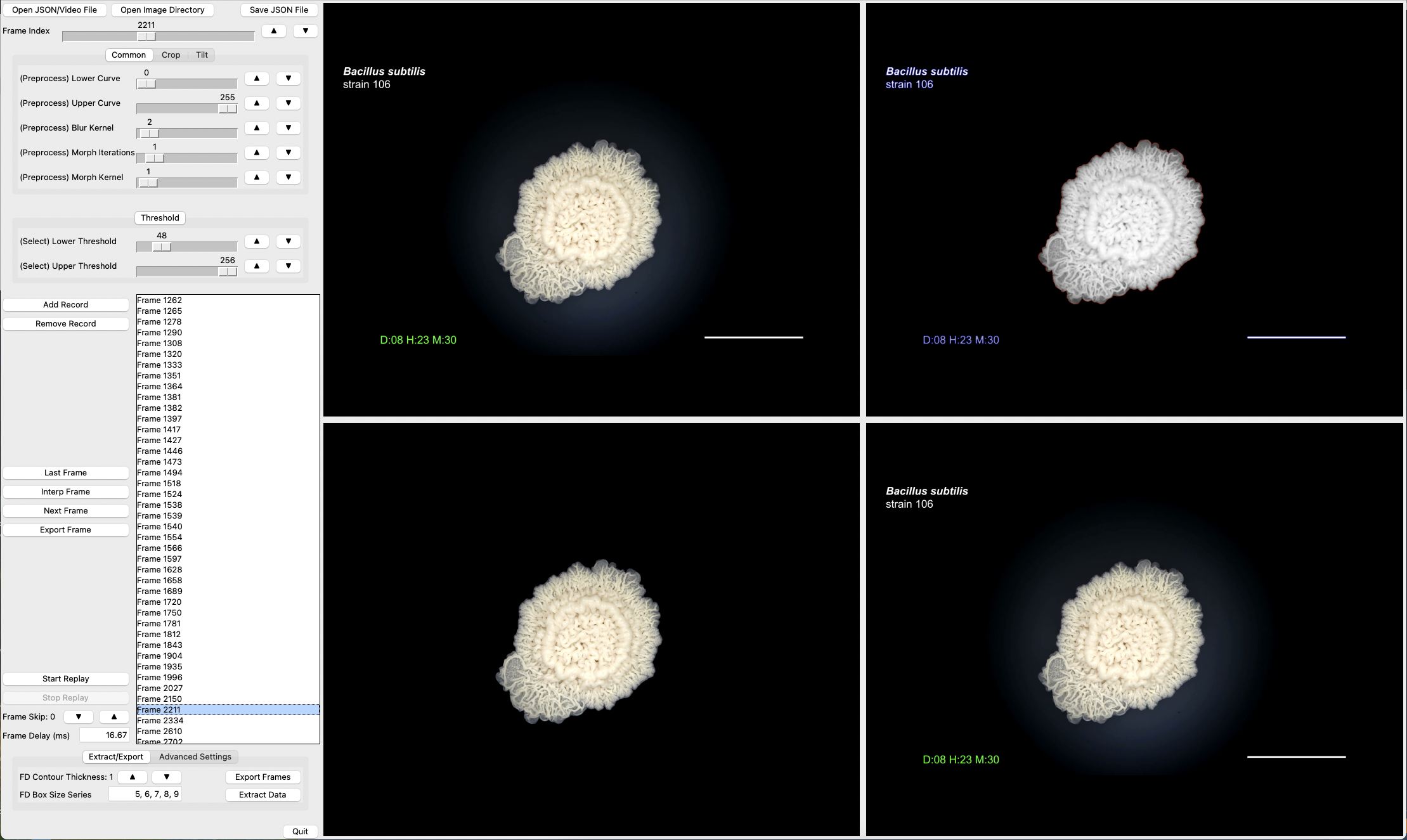}
    \caption{Graphical user interface (GUI) of \texttt{PyPETANA} during
    interactive segmentation validation and parameter selection as applied to strain 106 from Ref. \cite{Futo2022_SciRep,himberg2026geometrydynamicalmorphologygrowing}. The interface displays the raw input frame
    (top left), the corresponding binary mask and extracted contour overlay
    (top right), together with auxiliary diagnostic views used to assess
    preprocessing and threshold choices. Slider controls and toggles expose
    segmentation and analysis parameters, which are stored as immutable
    parameter snapshots (\texttt{records}) and subsequently applied during
    batch execution. The GUI acts as an orchestration and validation layer,
    while all numerical computations are performed by backend routines
    independent of the interface.}
    \label{fig:gui}
\end{figure}

Figure~\ref{fig:gui} illustrates the \texttt{PyPETANA} GUI during
interactive use. The interface is designed to support human-in-the-loop
validation of preprocessing and segmentation choices prior to batch analysis.
Users adjust parameters through sliders and toggles while inspecting
real-time overlays of binary masks and extracted contours on representative
frames. Once validated, parameter values are stored as explicit snapshot
records and remain fixed during batch execution. All subsequent numerical
analysis is therefore performed using frozen parameter sets, preventing
adaptive or time-dependent modification of segmentation settings during
processing.

A central design principle of \texttt{PyPETANA} is the strict separation
between the GUI and the numerical backend. The GUI acts exclusively as an
orchestration and validation layer, forwarding user-selected parameters to
backend analysis routines. The following section describes the overall
architecture of the codebase and the organization of the analysis pipeline.


\section{Codebase overview}
\label{sec:overview}

\texttt{PyPETANA} is an open-source Python framework for extracting
time-resolved geometric observables from evolving morphologies captured
in videos or ordered image sequences. The framework operates uniformly
across all frames in a dataset, treating morphology as a measurable
geometric object whose evolution may be quantified directly from image
data. Analysis parameters are selected by the user to capture the
relevant geometric features of the observed patterns.

Rather than coupling image analysis to biological, agent-based, or
statistical growth models, \texttt{PyPETANA} provides a direct mapping
from image data to explicitly defined geometric observables. This design
allows morphology to be analyzed as a dynamical geometric signal,
facilitating reproducible comparison across experiments, parameter
regimes, and independent datasets.

\subsection{Design philosophy}
\label{sec:design_philosophy}

\texttt{PyPETANA} is built around three guiding principles:
(i) geometry-first analysis without microscopic or model-dependent
assumptions,
(ii) strict separation between image preprocessing and observable
definition, and
(iii) deterministic, time-resolved execution under explicit parameter
snapshots.

Accordingly, the framework does not embed growth models, agent-based
rules, machine-learning inference, or data-driven feature extraction.
All reported observables are defined explicitly in terms of binary masks
and contour geometry derived independently from each frame.

This separation ensures that variations in reported observables reflect
changes in morphology itself rather than adaptive analysis choices or
model-dependent assumptions. The same backend routines are used for
interactive preview, batch execution, and multiprocessing, ensuring
numerical consistency across execution modes.

\subsection{Computational stages of \texttt{PyPETANA}}
\label{sec:stages}

\texttt{PyPETANA} processes image data on a frame-by-frame basis through
a fixed sequence of computational stages. Each frame is analyzed
independently under a well-defined parameter set, with temporal
consistency enforced through explicit parameter snapshots
(\texttt{records}) stored within the workflow.

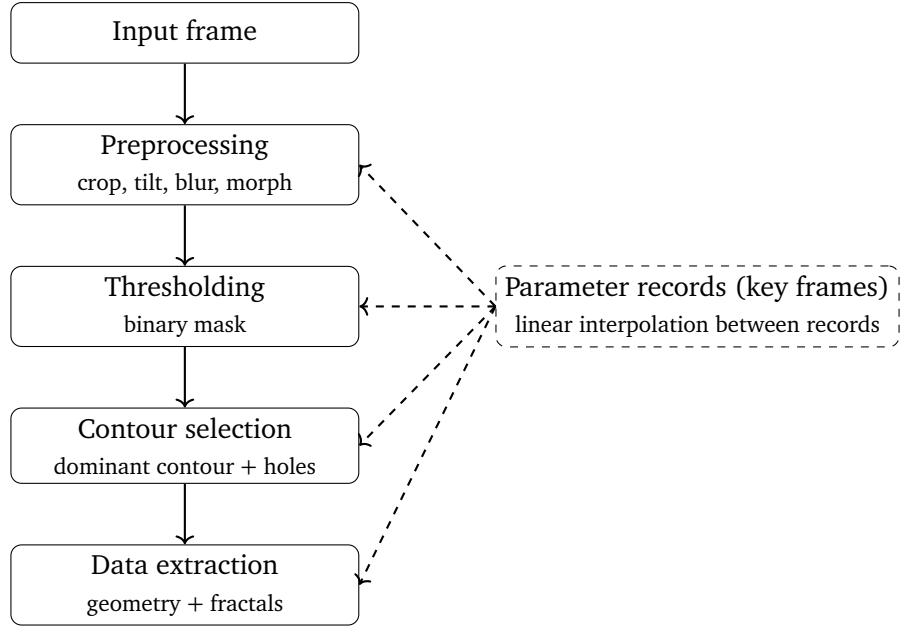
\begin{figure}[H]
\centering
\begin{tikzpicture}[
  node distance=8mm,
  stage/.style={draw, rounded corners, align=center, minimum width=4.6cm, minimum height=8mm},
  arrow/.style={->, thick}
]

\node[stage] (frame) {Input frame};
\node[stage, below=of frame] (prep) {Preprocessing\\\footnotesize crop, tilt, blur, morph};
\node[stage, below=of prep] (thresh) {Thresholding\\\footnotesize binary mask};
\node[stage, below=of thresh] (cont) {Contour selection\\\footnotesize dominant contour + holes};
\node[stage, below=of cont] (data) {Data extraction\\\footnotesize geometry + fractals};

\draw[arrow] (frame) -- (prep);
\draw[arrow] (prep) -- (thresh);
\draw[arrow] (thresh) -- (cont);
\draw[arrow] (cont) -- (data);

\node[draw, dashed, rounded corners, align=center, right=18mm of thresh, minimum width=4.6cm]
(records) {Parameter records (key frames)\\\footnotesize linear interpolation between records};

\draw[arrow, dashed] (records.west) -- (prep.east);
\draw[arrow, dashed] (records.west) -- (thresh.east);
\draw[arrow, dashed] (records.west) -- (cont.east);
\draw[arrow, dashed] (records.west) -- (data.east);

\end{tikzpicture}
\caption{Per-frame computational stages of \texttt{PyPETANA}.
Each frame is processed sequentially through preprocessing,
thresholding, contour selection, and data extraction.
Stage parameters are obtained from explicit parameter records
(key frames) and linearly interpolated for intermediate frames,
ensuring temporally smooth and reproducible execution.}
\label{fig:stages}
\end{figure}

For each frame, analysis proceeds through four stages:
preprocessing, thresholding, contour selection, and data extraction.
Together, these define the complete numerical pipeline implemented in
\texttt{PyPETANA}, illustrated schematically in
Fig.~\ref{fig:stages}.

\paragraph{Stage 1: Preprocessing.}
Each input frame is subjected to optional preprocessing operations
designed to improve segmentation robustness under non-uniform
illumination and imaging noise. These operations may include cropping,
luminance-tilt correction, contrast remapping, Gaussian blurring,
and morphological filtering. The output of this stage is a filtered
grayscale image optimized for segmentation.

\paragraph{Stage 2: Thresholding.}
Thresholding is applied to the preprocessed image to generate a binary
mask separating the morphology from the background. Threshold
parameters are user-defined and, once validated, are applied
consistently across frames through the parameter snapshot mechanism.

\paragraph{Stage 3: Contour selection.}
Contour detection is performed on the binary mask using OpenCV routines.
A dominant contour is selected using a geometric weighting criterion
that favors both large area and proximity to the image center.
Contours external to the dominant contour are discarded, while fully
enclosed contours may optionally be retained for hole-aware analysis.

\paragraph{Stage 4: Data extraction.}
Geometric observables, including area and perimeter, are computed
directly from the dominant contour. The contour is subsequently
rasterized to generate binary masks used for supersampled
box-counting analysis. Fractal dimensions of both the filled region
and a finite-width boundary band are computed by averaging over
multiple grid offsets to reduce alignment bias.

\paragraph{Parameter records and key frames.}
All preprocessing, thresholding, and analysis parameters are stored as
explicit parameter snapshots (\texttt{records}). Frames associated
with a record act as key frames and use the specified parameters
directly. For intermediate frames, parameter values are obtained
through linear interpolation between neighboring records. This
approach enables temporally smooth parameter variation while
preserving deterministic execution.


\section{I/O: video input, frame extraction, and time indexing}
\label{sec:io}

This section describes the \textbf{I/O core} of \texttt{PyPETANA},
which is responsible for media handling, frame extraction, and the
assignment of deterministic frame-based temporal indices to analyzed
images. The I/O core defines the interface between acquisition-specific
media formats and the frame-resolved analysis pipeline used throughout
the framework.

Once media have been decoded into image frames, all downstream analysis
stages operate exclusively on individual frames and their derived binary
masks. No subsequent computation depends on the original container
format or encoding scheme.

\subsection{Primary input: time-lapse video files (\texttt{.mov})}
\label{sec:input_mov}

The primary input to \texttt{PyPETANA} is a time-lapse video file
(e.g.\ \texttt{.mov}) containing an evolving morphology recorded over
time. When a video is loaded through the graphical user interface, the
I/O core decodes the media into an ordered sequence of image frames.
Each frame is treated as an independent discrete time point for all
subsequent analysis stages.

Internally, video decoding is performed using standard computer-vision
libraries~\cite{bradski2008learning} through deterministic,
index-based frame retrieval within the limits of the underlying video
backend. Frames are represented as in-memory image arrays and accessed
exclusively through a frame-indexed interface. Consequently, downstream
segmentation, geometry, fractal, and export routines operate solely on
decoded image frames and remain agnostic to the original media
container format, codec, or compression scheme.

As a result, the downstream analysis pipeline does not distinguish
between data originating from video files and data originating from
directories of pre-extracted images.

\subsection{Frame extraction and time indexing}
\label{sec:frame_extraction}

During input handling, frames are treated as a discrete ordered
sequence. Each frame is assigned a monotonically increasing integer
index $i$, which serves as the canonical temporal coordinate for all
time-resolved observables reported by \texttt{PyPETANA}. This
frame-indexed time coordinate is used internally throughout the
framework and preserved in all exported numerical outputs.

When acquisition metadata such as the frame rate (frames per second)
are available, the frame index may be mapped to physical time through

\[
t = \frac{i}{\mathrm{fps}},
\]

for purposes of interpretation and visualization. When such metadata
are unavailable or unreliable, analysis proceeds using the frame index
itself as the natural time variable. Throughout this article, time is
denoted by $t \equiv i$ unless stated otherwise.

The output of the I/O core is therefore a deterministic,
frame-indexed stream of image data that forms the exclusive input to
the segmentation stage described in Sec.~\ref{subsec:segmentation}.

\subsection{Image sequences}
\label{sec:image_sequences}

Users who already possess time-resolved image data in the form of
ordered image files (e.g.\ \texttt{.png}, \texttt{.jpg},
\texttt{.jpeg}, \texttt{.bmp}, or \texttt{.tiff}) may provide an image
directory instead of a video file. In this case, the I/O core
enumerates supported image files in lexicographically sorted order and
treats each image as a successive time point.

No other component of the analysis pipeline is modified. The
segmentation, geometry, fractal, and export stages operate identically
on frames originating from videos or image sequences, and all
observables are indexed using the same frame identifier. In all cases,
the I/O core produces a deterministic, frame-indexed image stream that
serves as the exclusive input to the segmentation pipeline.


\section {Main stages/pipeline of \texttt{PyPETANA}}

\subsection{Segmentation core: Preprocessing, mask generation \& thresholding}
\label{subsec:segmentation}

This section documents the segmentation core of \texttt{PyPETANA}, which
transforms raw image frames into binary masks representing the spatial footprint
of an evolving morphology. The segmentation core provides the sole interface between raw imaging data and all
subsequent geometric and fractal analyses performed by the codebase (Fig.~\ref{fig:segmentation_pipeline}). 

\subsubsection{Segmentation pipeline}
\label{sec:segmentation_pipeline}
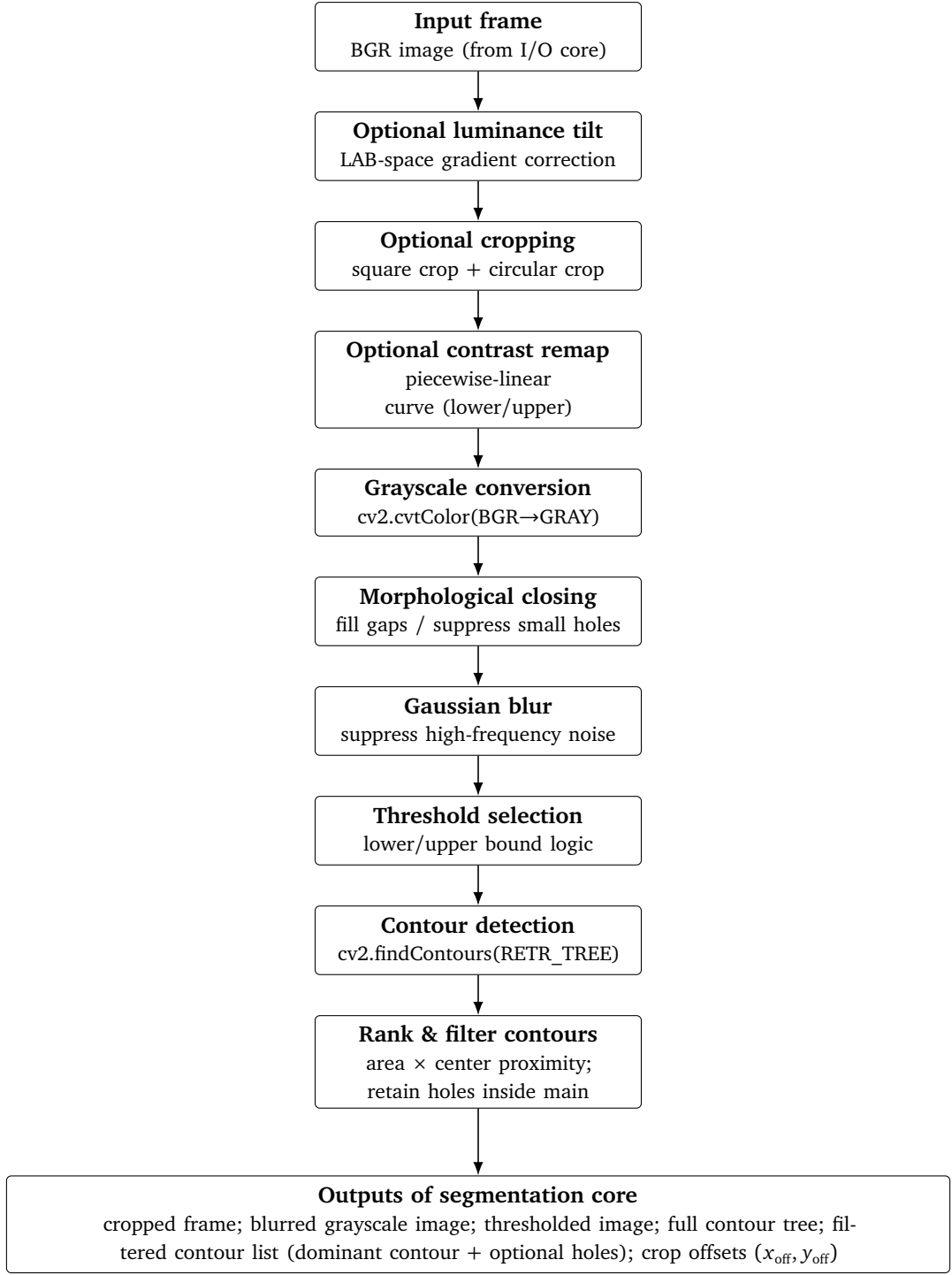
\begin{figure}[H]
\centering
\begin{tikzpicture}[node distance=6mm and 10mm]
  \node[pnode] (in) {\textbf{Input frame}\\ \footnotesize BGR image (from I/O core)};
  \node[pnode, below=of in] (tilt) {\textbf{Optional luminance tilt}\\ \footnotesize LAB-space gradient correction};
  \node[pnode, below=of tilt] (crop) {\textbf{Optional cropping}\\ \footnotesize square crop + circular crop};
  \node[pnode, below=of crop] (curve) {\textbf{Optional contrast remap}\\ \footnotesize piecewise-linear curve (lower/upper)};
  \node[pnode, below=of curve] (gray) {\textbf{Grayscale conversion}\\ \footnotesize cv2.cvtColor(BGR$\rightarrow$GRAY)};
  \node[pnode, below=of gray] (morph) {\textbf{Morphological closing}\\ \footnotesize fill gaps / suppress small holes};
  \node[pnode, below=of morph] (blur) {\textbf{Gaussian blur}\\ \footnotesize suppress high-frequency noise};
  \node[pnode, below=of blur] (thr) {\textbf{Threshold selection}\\ \footnotesize lower/upper bound logic};
  \node[pnode, below=of thr] (cont) {\textbf{Contour detection}\\ \footnotesize cv2.findContours(RETR\_TREE)};
  \node[pnode, below=of cont] (rank) {\textbf{Rank \& filter contours}\\ \footnotesize area $\times$ center proximity; retain holes inside main};

  \draw[parrow] (in) -- (tilt);
  \draw[parrow] (tilt) -- (crop);
  \draw[parrow] (crop) -- (curve);
  \draw[parrow] (curve) -- (gray);
  \draw[parrow] (gray) -- (morph);
  \draw[parrow] (morph) -- (blur);
  \draw[parrow] (blur) -- (thr);
  \draw[parrow] (thr) -- (cont);
  \draw[parrow] (cont) -- (rank);

  \node[pwide, below=10mm of rank] (out) {\textbf{Outputs of segmentation core}\\
  \footnotesize cropped frame; blurred grayscale image; thresholded image; full contour tree;
  filtered contour list (dominant contour + optional holes); crop offsets $(x_{\mathrm{off}},y_{\mathrm{off}})$};
  \draw[parrow] (rank) -- (out);
\end{tikzpicture}
\caption{Segmentation pipeline implemented by \lstinline{preprocess\_frame} (see algorithm~\ref{alg:preprocess_frame}.)
Optional steps (luminance tilt, cropping, contrast remapping) are controlled by GUI parameters.
The primary outputs are the thresholded image and the filtered contour set that
serve as inputs to \lstinline{export\_frame}  and \lstinline{extract\_data}(see algorithm~\ref{alg:extract_data}). }
\label{fig:segmentation_pipeline}
\end{figure}

The segmentation core converts each raw frame into a binary mask through a
deterministic, geometry-consistent pipeline.
All operations are applied independently to each frame under a fixed parameter
snapshot (record), ensuring frame-wise reproducibility and eliminating
time-dependent parameter drift.
Figure~\ref{fig:segmentation_pipeline} provides a schematic overview of the
segmentation pipeline implemented by
\lstinline{preprocess_frame} (see Algorithm ~\ref{alg:preprocess_frame}). Optional preprocessing steps are controlled through the GUI but, once validated,
are held fixed for the duration of a batch analysis.
The outputs of the segmentation core include intermediate grayscale images,
thresholded masks, and the complete contour tree, which together form the input to
subsequent contour selection and data extraction stages.

\subsubsection{Role of segmentation in the analysis pipeline}
The primary purpose of the segmentation core is to produce a binary geometric
representation of the morphology at each time point, separating the region of
interest from the background.
All downstream computations—including area, perimeter, circularity, and fractal
dimension estimation—operate exclusively on these binary masks and their derived
contours.
As a result, segmentation is the only stage of the pipeline that directly depends
on imaging conditions such as illumination, contrast, and noise.
Importantly, \texttt{PyPETANA} treats segmentation as a purely geometric
abstraction.
The segmentation core does not assume any specific biological growth mechanism,
cellular structure, or microscopic dynamics and remains agnostic to the physical
origin of the observed patterns.

\subsubsection{Preprocessing}
\label{sec:preprocessing}
Each input frame is subjected to a fixed sequence of optional preprocessing
operations designed to improve segmentation robustness under non-uniform
illumination and imaging noise (Algorithm~\ref{alg:preprocess_frame}).
These operations correspond directly to the early stages of the pipeline shown in
Fig.~\ref{fig:segmentation_pipeline} and include luminance-tilt correction,
cropping, contrast remapping, grayscale conversion, morphological closing, and
Gaussian smoothing.
All preprocessing parameters are selected interactively through the GUI and
validated by visual inspection.
Once validated, the parameter snapshot is applied uniformly to every frame in a
batch analysis.
This guarantees temporal consistency and prevents frame-to-frame adaptation of
segmentation parameters.

\subsubsection{Binary mask generation and thresholding}
\label{sec:binary_mask}
Following preprocessing, each frame is converted into a binary mask using
threshold-based segmentation.
Pixels belonging to the morphology are assigned a value of one, while background
pixels are assigned zero.
Thresholds may be specified as lower bounds, upper bounds, or combined intensity
intervals, as detailed in Algorithm~\ref{alg:preprocess_frame}.
Threshold parameters are fixed within a given parameter record and applied
consistently across all frames associated with that record.
Consequently, time-dependent trends in extracted observables reflect changes in
morphology rather than adaptive segmentation choices.
The segmentation core produces exactly one binary mask per frame.
All subsequent observables are computed from contour geometry and derived masks,
rather than from raw pixel intensities.

\subsubsection{Interactive validation via the GUI}

\begin{figure}[H]
\centering
\includegraphics[width=0.9\linewidth]{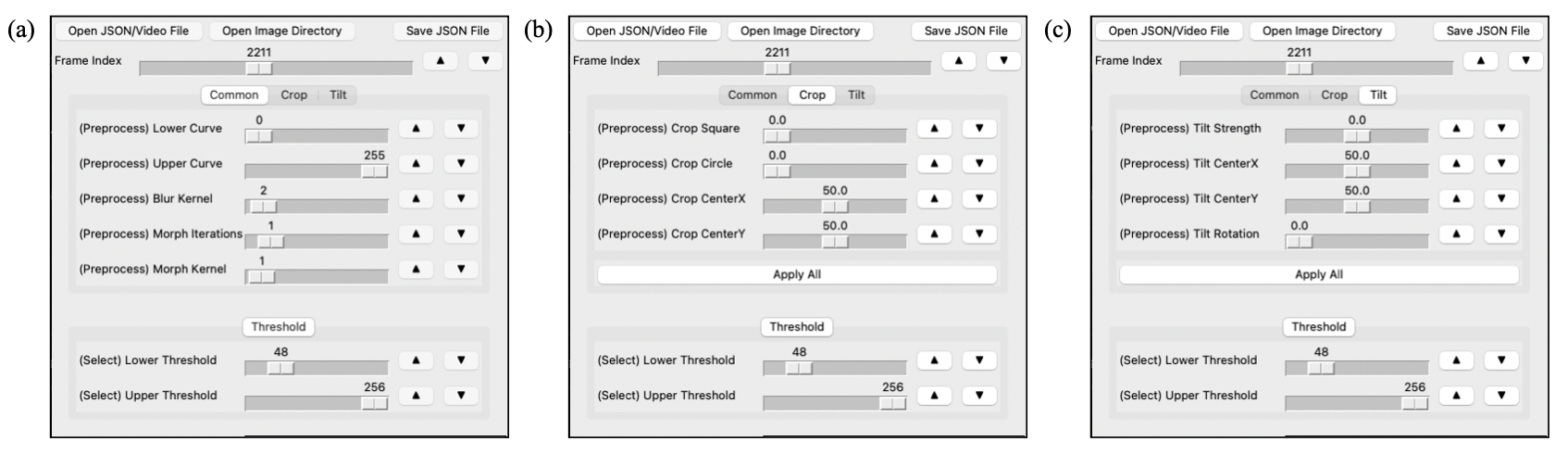}
\caption{
Graphical interface used to define segmentation parameters prior to batch execution.
(a) Common preprocessing parameters, including contrast remapping,
Gaussian blur kernel, and morphological closing.
(b) Cropping controls (square and circular).
(c) Luminance-tilt correction parameters.
Threshold bounds are specified separately.
After validation, parameter values are stored in a record and applied
uniformly across frames, ensuring deterministic and reproducible segmentation.
}
\label{fig:gui_segmentation}
\end{figure}

A key feature of \texttt{PyPETANA} is the ability to visually validate segmentation
results prior to executing full batch analyses.
Through the GUI, users can preview raw frames alongside their corresponding binary
masks and detected contours and adjust segmentation parameters as needed.
Once validated, parameter values are stored in an explicit record and applied
uniformly to all frames associated with that record.
This guarantees deterministic execution and ensures that time-resolved observables
reflect morphological evolution rather than frame-dependent parameter adaptation.
Contour ranking, introduced in the next subsection, is used solely as a geometric
disambiguation mechanism and does not encode any biological or dynamical
assumptions.

\begin{algorithm}[H]
\caption{\texttt{preprocess\_frame}: preprocessing, thresholding, and contour selection}
\label{alg:preprocess_frame}
\KwIn{
BGR frame $I$;\;
GUI parameters controlling luminance tilt, square/circular crop, contrast curve,
morphological closing, Gaussian blur, and threshold selection
}
\KwOut{
Cropped frame $I_{\mathrm{crop}}$;\;
blurred grayscale image $B$;\;
thresholded image $T$;\;
all detected contours $\mathcal{C}$;\;
filtered contour set $\mathcal{C}_{\star}$ (dominant contour + optional holes);\;
crop offsets $(x_{\mathrm{off}},y_{\mathrm{off}})$
}

\tcp{1. Optional luminance-tilt correction}
\If{tilt enabled}{
    Convert $I$ to LAB color space\;
    Modify L-channel using planar or linear gradient\;
    Convert back to BGR and update $I$\;
}

\tcp{2. Optional cropping}
\If{square crop enabled}{
    Crop around user-defined center; update offsets $(x_{\mathrm{off}},y_{\mathrm{off}})$\;
}
\If{circular crop enabled}{
    Apply circular mask; trim to bounding rectangle; update offsets\;
}

\tcp{3. Optional contrast remapping}
\If{contrast curve enabled}{
    Apply piecewise-linear intensity remapping\;
}

\tcp{4. Grayscale conversion, morphology, and smoothing}
$G \leftarrow \mathrm{cvtColor}(I;\mathrm{BGR}\rightarrow\mathrm{GRAY})$\;
$G_c \leftarrow \mathrm{morphClose}(G;\text{elliptic kernel, iterations})$\;
$B \leftarrow \mathrm{GaussianBlur}(G_c;\text{kernel size})$\;

\tcp{5. Threshold selection}
\uIf{both lower and upper thresholds active}{
    $M \leftarrow (B \ge \ell_T)\land(B < u_T)$\;
}
\uElseIf{only lower threshold active}{
    $M \leftarrow (B \ge \ell_T)$\;
}
\uElseIf{only upper threshold active}{
    $M \leftarrow (B < u_T)$\;
}
\Else{
    $M \leftarrow \text{all pixels}$\;
}
$T \leftarrow B \odot M$\;

\tcp{6. Contour detection and geometric filtering}
$\mathcal{C} \leftarrow \mathrm{findContours}(T;\mathrm{RETR\_TREE})$\;
Score each contour by (area $\times$ center-proximity weight)\;
Select highest-scoring contour as $c_0$\;
$\mathcal{C}_{\star} \leftarrow [c_0]$\;
\ForEach{contour $c \in \mathcal{C} \setminus \{c_0\}$}{
    \If{$c$ lies fully inside $c_0$}{
        Append $c$ to $\mathcal{C}_{\star}$ \tcp*{treated as interior hole}
    }
}
\Return $(I_{\mathrm{crop}},B,T,\mathcal{C},\mathcal{C}_{\star},
x_{\mathrm{off}},y_{\mathrm{off}})$\;
\end{algorithm}

The filtered contour set produced by the segmentation core serves as the exclusive
input to the observables core, where all geometric and fractal quantities are computed. 

\subsection{Contour selection}
\label{subsec:contour-sel}

Following binary mask generation, \texttt{PyPETANA} performs contour detection
using OpenCV’s hierarchical retrieval mode
(\lstinline{cv2.findContours} with \lstinline{RETR_TREE}).
This produces a complete contour tree containing all closed boundaries present
in the thresholded image, including both exterior objects and interior holes.

The purpose of contour selection is to identify a single dominant contour that
represents the morphology of interest in each frame, while retaining any
fully enclosed interior contours for optional hole-aware analysis.
This selection step is purely geometric and does not encode biological
or mechanistic assumptions.

\subsubsection{Moment-based centroid and proximity weighting}

For each detected contour $c_i$, \texttt{PyPETANA} computes its area
$A_i = \mathrm{area}(c_i)$ using \lstinline{cv2.contourArea}.
The contour centroid is obtained from spatial moments computed via
\lstinline{cv2.moments}.
Let $m_{pq}$ denote the $(p,q)$ moment returned by OpenCV.
When $m_{00}\neq 0$, the centroid coordinates are
\begin{equation}
(x_i,y_i)=\left(\frac{m_{10}}{m_{00}},\,\frac{m_{01}}{m_{00}}\right),
\end{equation}
and if $m_{00}=0$ the contour is assigned zero weight (degenerate case).

To suppress spurious peripheral contours and to regularize contour identity
across successive frames, contours are ranked by a score that combines
area with a smooth proximity weight relative to a reference center
$\mathbf{x}_{\mathrm{ref}}=(x_{\mathrm{ref}},y_{\mathrm{ref}})$.
In the implementation, $\mathbf{x}_{\mathrm{ref}}$ is the center of the
current frame after cropping has been applied. Rather than tracking the center
automatically this allows the user to select the center, adjusting
key-frames to track an area of interest dynamically.

Defining the squared displacement
\begin{equation}
d_i^2 = (x_i-x_{\mathrm{ref}})^2 + (y_i-y_{\mathrm{ref}})^2,
\end{equation}
the proximity weight is chosen as the bounded rational decay
\begin{equation}
w(d_i) = \frac{1}{1+d_i^2},
\end{equation}
leading to the contour score
\begin{equation}
S_i = A_i\, w(d_i)
    = \frac{A_i}{1+(x_i-x_{\mathrm{ref}})^2+(y_i-y_{\mathrm{ref}})^2}.
\end{equation}

The rational decay $w(d_i)$ is strictly positive, smooth, and monotonically
decreasing, providing a soft proximity prior without introducing hard spatial
cutoffs.
Consequently, contour selection remains primarily geometry-driven
(area-dominated) while being stabilized against frame-to-frame identity
switches in the presence of multiple candidates.

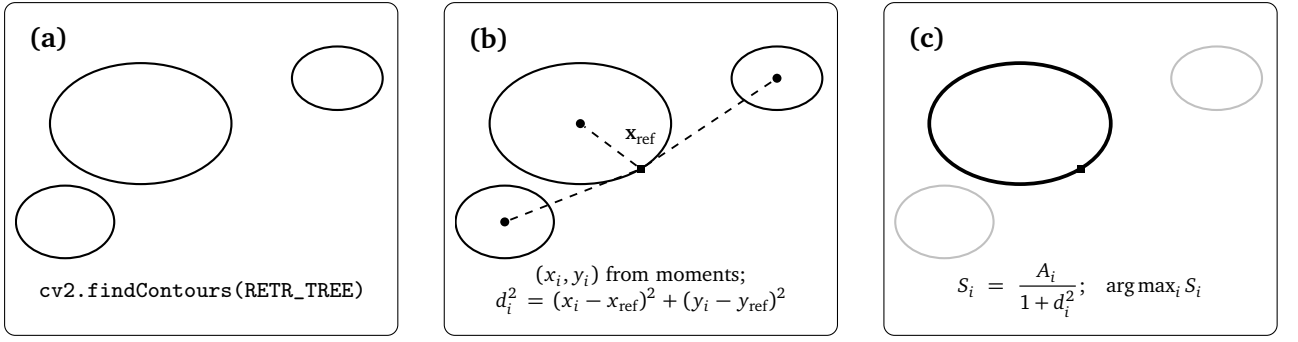
\begin{figure}[H]
\centering
\begin{tikzpicture}[
  panel/.style={draw, rounded corners, minimum width=5.2cm, minimum height=4.4cm},
  lab/.style={font=\bfseries},
  centroid/.style={fill=black, circle, inner sep=1.2pt},
  refcenter/.style={fill=black, rectangle, inner sep=1.4pt},
  contour/.style={draw, line width=0.8pt},
  selected/.style={draw, line width=1.4pt},
  hole/.style={draw, dashed, line width=0.9pt},
  dist/.style={dashed, line width=0.7pt},
  strip/.style={font=\scriptsize, align=center, text width=4.9cm}
]

\def\stripy{-16mm}

\def\dx{-8mm}\def\dy{6mm}
\def\sx{18mm}\def\sy{12mm}
\def\tx{-18mm}\def\ty{-7mm}

\node[panel] (A) {};
\node[lab, anchor=north west] at ([xshift=2mm,yshift=-2mm]A.north west) {(a)};

\begin{scope}
  \clip ([xshift=1.5mm,yshift=10mm]A.south west)
        rectangle
        ([xshift=-1.5mm,yshift=-4mm]A.north east);

  \draw[contour] ([xshift=\dx,yshift=\dy]A.center) ellipse (12mm and 8mm);
  \draw[contour] ([xshift=\sx,yshift=\sy]A.center) ellipse (6mm and 4.2mm);
  \draw[contour] ([xshift=\tx,yshift=\ty]A.center) ellipse (6.5mm and 4.8mm);
\end{scope}

\node[strip] at ([yshift=\stripy]A.center)
{\texttt{cv2.findContours(RETR\_TREE)}};

\node[panel, right=6mm of A] (B) {};
\node[lab, anchor=north west] at ([xshift=2mm,yshift=-2mm]B.north west) {(b)};

\begin{scope}
  \clip ([xshift=1.5mm,yshift=10mm]B.south west)
        rectangle
        ([xshift=-1.5mm,yshift=-4mm]B.north east);

  \draw[contour] ([xshift=\dx,yshift=\dy]B.center) ellipse (12mm and 8mm);
  \draw[contour] ([xshift=\sx,yshift=\sy]B.center) ellipse (6mm and 4.2mm);
  \draw[contour] ([xshift=\tx,yshift=\ty]B.center) ellipse (6.5mm and 4.8mm);

  \node[refcenter] (ref) at (B.center) {};
  \node[font=\scriptsize, anchor=south] at ([yshift=2mm]ref)
    {$\mathbf{x}_{\mathrm{ref}}$};

  \node[centroid] (c1) at ([xshift=\dx,yshift=\dy]B.center) {};
  \node[centroid] (c2) at ([xshift=\sx,yshift=\sy]B.center) {};
  \node[centroid] (c3) at ([xshift=\tx,yshift=\ty]B.center) {};

  \draw[dist] (ref) -- (c1);
  \draw[dist] (ref) -- (c2);
  \draw[dist] (ref) -- (c3);
\end{scope}

\node[strip] at ([yshift=\stripy]B.center)
{$(x_i,y_i)$ from moments;\quad
$d_i^2=(x_i-x_{\mathrm{ref}})^2+(y_i-y_{\mathrm{ref}})^2$};

\node[panel, right=6mm of B] (C) {};
\node[lab, anchor=north west] at ([xshift=2mm,yshift=-2mm]C.north west) {(c)};

\begin{scope}
  \clip ([xshift=1.5mm,yshift=10mm]C.south west)
        rectangle
        ([xshift=-1.5mm,yshift=-4mm]C.north east);

  \draw[selected] ([xshift=\dx,yshift=\dy]C.center) ellipse (12mm and 8mm);


  \draw[contour, opacity=0.25]
    ([xshift=\sx,yshift=\sy]C.center) ellipse (6mm and 4.2mm);

  \draw[contour, opacity=0.25]
    ([xshift=\tx,yshift=\ty]C.center) ellipse (6.5mm and 4.8mm);

  \node[refcenter] at (C.center) {};
\end{scope}

\node[strip] at ([yshift=\stripy]C.center)
{$S_i=\dfrac{A_i}{1+d_i^2}$;\quad
$\operatorname*{arg\,max}_i S_i$};

\end{tikzpicture}
\caption{
Moment-based contour selection as implemented in \texttt{PyPETANA}.
(a) Multiple contours detected after thresholding using \texttt{cv2.findContours}.
(b) Spatial moments (\texttt{cv2.moments}) provide centroid coordinates;
distances are measured relative to the reference center
$\mathbf{x}_{\mathrm{ref}}$ (center of the current frame after cropping is applied).
(c) Contours are ranked using $S_i=A_i/(1+d_i^2)$ and the highest-scoring
contour is selected as the dominant morphology.
Here $(x_i,y_i)$ denote centroids computed from spatial moments and
$\mathbf{x}_{\mathrm{ref}}$ is the reference center used for proximity weighting.
}
\label{fig:contour_selection_schematic}
\end{figure}
\subsubsection{Hole handling and contour filtering}

After selecting the dominant contour $c_0$, the contour hierarchy returned
by \lstinline{RETR_TREE} is used to identify contours fully contained within
$c_0$.
Contours that lie entirely inside $c_0$ are retained as interior contours
(holes), while all other contours are discarded.

The filtered contour set
\begin{equation}
\mathcal{C}_{\star}
= \{c_0\} \cup
  \{\text{interior contours of } c_0\}
\end{equation}
is passed to the observables core for geometric and fractal analysis.

This procedure ensures that exactly one primary morphology is extracted per
frame while preserving interior topology when required.
Because selection is based solely on contour geometry and image moments,
the method remains deterministic, translation-consistent, and independent
of domain-specific assumptions.


\subsection{Data extraction of physical observables}
\label{subsec:observables}

This section documents how \texttt{PyPETANA} converts the filtered contour set
$\mathcal{C}_\star$ returned by segmentation into quantitative, frame-resolved
observables. All quantities are computed independently for each frame index and
depend exclusively on the binary mask geometry induced by the selected contour(s).
Raw pixel intensities influence the analysis only indirectly through segmentation;
they do not enter directly into observable definitions.

The observables implemented in \texttt{PyPETANA} are organized into two complementary
classes:
(i) \emph{geometric observables} that quantify global size and compactness, and
(ii) \emph{fractal observables} that probe multiscale spatial complexity of both the
filled morphology (mass scaling) and its finite-width boundary band (roughness scaling).
Together, these descriptors provide a deterministic morphological characterization
without invoking biological, statistical, or mechanistic models.

In the reference implementation, per-frame computations are performed by the routine
\lstinline{extract_data} (Algorithm~\ref{alg:extract_data}). By default, observables
are computed for the dominant connected component (the \emph{main} contour).
An optional flag (\lstinline{include_holes}) enables hole-adjusted variants when
interior perforations are physically meaningful and reliably segmented.

\subsubsection{Inputs and outputs}
\label{sec:observables_io}

For each analyzed frame, \lstinline{extract_data} takes:
(i) a preprocessed image array \texttt{blur} (used only to fix mask dimensions),
(ii) the filtered contour list $\mathcal{C}_\star=[c_0,c_1,\dots]$, and
(iii) analysis parameters controlling boundary thickness and box-size sampling.

The primary outputs (one row per frame) are:
\begin{itemize}
  \item \textbf{Geometric observables:} area $A$, perimeter $P$, and circularity $C$.
  \item \textbf{Fractal observables:} filled-mask fractal dimension $D_f$ (mass scaling)
        and boundary-band fractal dimension $D_b$ (roughness scaling),
        together with RMS residuals from the corresponding log--log fits as
        diagnostics of scaling consistency.
\end{itemize}

If \lstinline{include_holes} is enabled and interior contours are present,
additional hole-adjusted observables are returned alongside the primary metrics.

\subsubsection{Geometric observables}
\label{sec:geometry}

Let $\mathcal{C}_\star=[c_0,c_1,\dots]$ denote the filtered contour list for a frame,
where $c_0$ is the dominant external contour and any $c_{i\ge 1}$ are fully enclosed
interior contours (holes), when present.

For the dominant contour, \texttt{PyPETANA} computes:
\begin{equation}
A = \mathrm{Area}(c_0), \qquad
P = \mathrm{Perimeter}(c_0),
\end{equation}
using OpenCV’s polygonal contour geometry
(\lstinline{cv2.contourArea}, \lstinline{cv2.arcLength} with closed contour flag).
Circularity is defined as
\begin{equation}
C=\frac{4\pi A}{P^2},
\end{equation}
which equals $1$ for a perfect circle and decreases as shapes become less compact
or more irregular.

These geometric observables capture global morphology (size and compactness)
and are intentionally less sensitive to fine-scale boundary roughness than the
fractal descriptors introduced below.

\paragraph{Optional: hole-adjusted geometry.}
If \lstinline{include_holes=True} and interior contours exist, total hole area
and perimeter are accumulated as
\begin{equation}
A_h=\sum_{k\ge 1}\mathrm{Area}(c_k), \qquad
P_h=\sum_{k\ge 1}\mathrm{Perimeter}(c_k).
\end{equation}
A hole-adjusted circularity is then defined as
\begin{equation}
C_{\mathrm{holes}}=
\frac{4\pi\left(A-A_h\right)}{\left(P+P_h\right)^2}.
\end{equation}

Hole-adjusted metrics are meaningful when interior voids are stable and
physically relevant. For many datasets, the dominant external boundary alone
provides a robust morphological descriptor; accordingly, hole-related metrics
are disabled by default and activated only at user discretion.

\subsubsection{Fractal observables via supersampled box counting}
\label{sec:fractal}

While geometric observables summarize global shape, fractal observables
characterize multiscale spatial complexity. \texttt{PyPETANA} computes
effective box-counting dimensions over a controlled range of length scales
using a deterministic supersampled box-counting procedure.

\paragraph{Mask construction.}
The dominant contour $c_0$ is rasterized into a filled binary mask
$M\in\{0,1\}^{H\times W}$ using
\lstinline{cv2.drawContours} with \lstinline{thickness=cv2.FILLED}.
The filled-mask fractal dimension $D_f$ quantifies \emph{mass scaling}, i.e.,
how the occupied area scales with box size.

To isolate boundary roughness at a finite thickness, an inner boundary band
$M_{\partial}$ is constructed via morphological erosion:
\begin{equation}
E=\mathrm{erode}(M;\tau), \qquad
M_{\partial}= M \wedge (1-E),
\end{equation}
where $\tau$ (\lstinline{contour_thickness}) denotes the number of erosion
iterations using a fixed elliptical structuring element.
The boundary-band fractal dimension $D_b$ thus quantifies
\emph{roughness scaling} of the morphology boundary rather than bulk mass scaling.

If \lstinline{include_holes=True}, interior contours are filled with zero
to produce a holes-removed mask $M^{(h)}$ and corresponding boundary band
$M^{(h)}_{\partial}$.
For numerical stability, masks are cropped to their nonzero bounding box
prior to box counting; this cropping removes empty background without
altering geometry.

\paragraph{Supersampled box counting with alignment averaging.}
For each box size $b$, masks are zero-padded so that both dimensions are divisible by $b$.
To reduce grid-alignment bias, four distinct padding placements per $b$
(top/left shifts) are evaluated.
For each placement, the padded mask is reshaped into non-overlapping
$b\times b$ blocks and the number of occupied boxes is computed:
\begin{equation}
N(b)=\#\{\text{blocks with nonzero pixel sum}\}.
\end{equation}

This is performed separately for the filled mask ($N_f(b)$) and boundary band
($N_b(b)$), and likewise for hole-adjusted masks when enabled.

\paragraph{Log--log fits and effective dimensions.}
Effective fractal dimensions are extracted from linear least-squares fits in
log--log space:
\begin{equation}
\log N(b)= -D\,\log b + \mathrm{const}.
\end{equation}
With this convention, the fitted slope is negative and the reported dimension
is $D=-\mathrm{slope}$.
\texttt{PyPETANA} reports:
\begin{itemize}
  \item $D_f$ — effective fractal dimension of the filled region (mass scaling),
  \item $D_b$ — effective fractal dimension of the boundary band (roughness scaling),
  \item RMS residuals of the fits as diagnostics of scaling consistency.
\end{itemize}

The use of multiple padding placements ensures that the extracted dimensions
are not artifacts of grid alignment, improving robustness across frames.

\begin{figure}[H]
    \centering
    \includegraphics[width=0.9\linewidth]{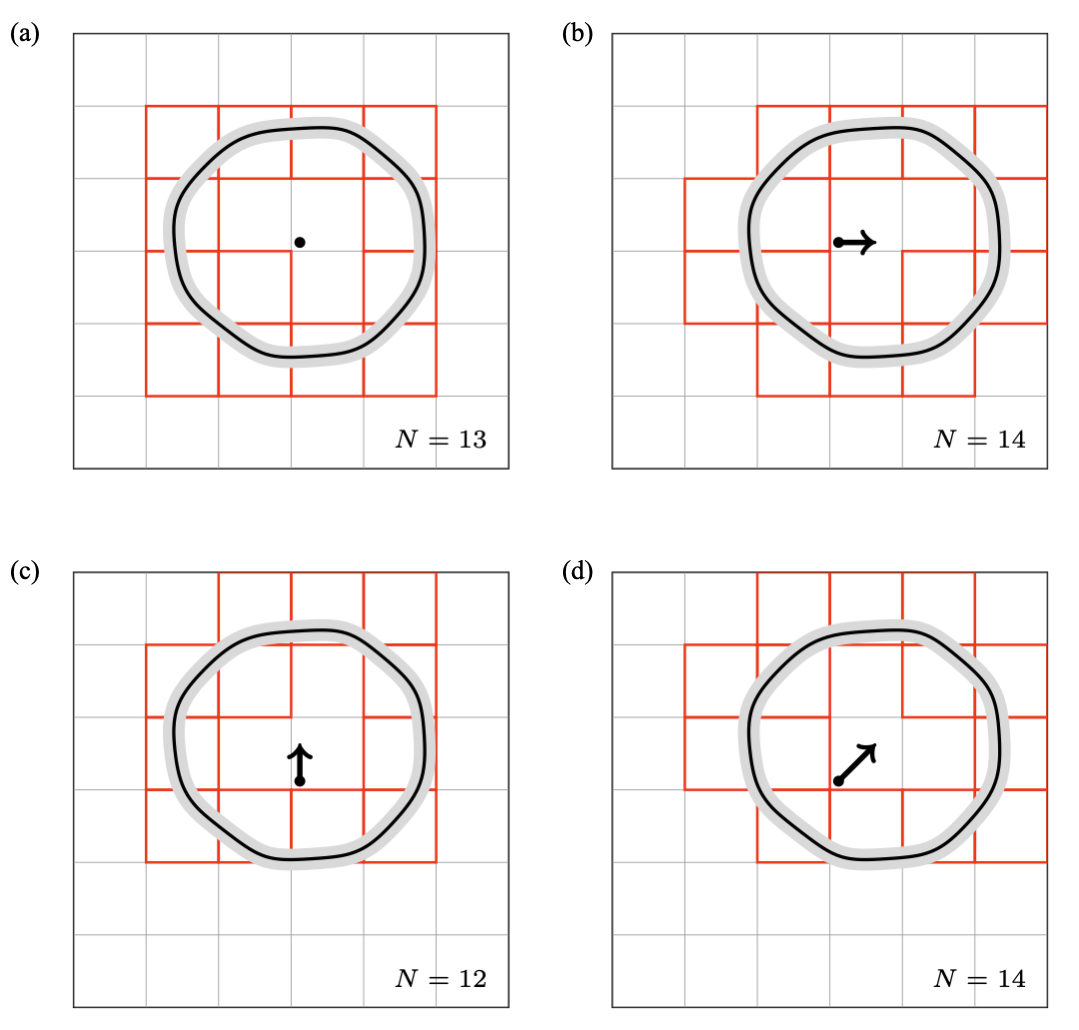}
    \caption{(a) are counted, this is the first and centered panel. The contour is then shifted half a box width in x, boxes intersected counted for panel (b). From center the contour is again shifted half a box width in y, boxes intersected counted for panel (c). Finally from center the contour is again shifted half a box width in y, and also in x $(\sqrt(x^2+y^2))$, boxes intersected for panel (d). }
    \label{fig:placeholder}
\end{figure}

\subsubsection{Algorithmic implementation}
\label{sec:extract_data_algo}

Algorithm~\ref{alg:extract_data} summarizes the per-frame computation of
geometric and fractal observables as implemented in
\lstinline{extract_data}. The procedure is invoked independently for each frame
and is fully determined by $\mathcal{C}_\star$ and the fixed analysis parameters.

\begin{algorithm}[H]
\caption{\texttt{extract\_data}: per-frame geometry and supersampled box-counting dimensions}
\label{alg:extract_data}
\KwIn{
Preprocessed image \texttt{blur} (mask shape only);\;
filtered contours $\mathcal{C}_\star=[c_0,c_1,\dots]$;\;
box-size seed string \texttt{box\_sizes\_series};\;
boundary thickness $\tau$;\;
flag \texttt{include\_holes}
}
\KwOut{
$A,P,C$;\;
$D_f, D_f^{\mathrm{RMS}}$;\;
$D_b, D_b^{\mathrm{RMS}}$;\;
(optional) hole-adjusted variants
}

\tcp{1. Geometry from dominant contour}
Compute $A$, $P$, and $C$\;

\If{\texttt{include\_holes} and $|\mathcal{C}_\star|>1$}{
    Compute $A_h$, $P_h$, and $C_{\mathrm{holes}}$\;
}

\tcp{2. Mask construction}
Rasterize $c_0$ into filled mask $M$\;
Construct boundary band $M_{\partial}$ via erosion with thickness $\tau$\;
Optionally construct hole-adjusted masks\;
Crop masks to nonzero bounding box\;

\tcp{3. Supersampled box counting}
Generate box sizes by expanding seeds by powers of 2\;
\ForEach{$b$}{
    For four padding placements:\;
    Zero-pad masks, reshape into $b\times b$ blocks, count occupied boxes\;
    Store $(\log b,\log N_f,\log N_b)$\;
}

\tcp{4. Log--log fits}
Fit $\log N_f(b)$ and $\log N_b(b)$ vs.\ $\log b$\;
Set $D_f=-\mathrm{slope}_f$, $D_b=-\mathrm{slope}_b$\;
Return observables\;
\end{algorithm}

\section{Export and visualization}
\label{sec:export}

The export and visualization core of \texttt{PyPETANA} writes analysis products to disk in a form that supports
(i) downstream time-series analysis,
(ii) visual validation of segmentation and contour selection, and
(iii) computational reproducibility.
All exported quantities are indexed by a frame identifier (and therefore by time; Sec.~\ref{sec:io}),
so that numerical observables, diagnostic plots, and rendered overlays remain cross-referencable on a per-frame basis.

\subsection{Output directory layout and naming conventions}
\label{sec:export_layout}

A \texttt{PyPETANA} run can write numerical outputs, GUI snapshots, exported frames, and fractal-dimension (FD)
diagnostics using a deterministic, frame-indexed naming scheme. In the current implementation:

\begin{itemize}
    \item \texttt{<chosen\_output>.dat} \hfill (primary numerical output table; Sec.~\ref{sec:export_numeric})
    \item \texttt{<chosen\_output>.json} \hfill (GUI parameter snapshots; replay/interpolation input; Sec.~\ref{sec:export_json})
    \item \texttt{<basename(media\_path)>.export/} \hfill (optional exported frames / overlays for validation)
    \item \texttt{<media\_path>.extract/} \hfill (optional FD fit tables and FD occupancy visualizations)
\end{itemize}

\paragraph{User-selectable output names.}
The primary numerical output file is chosen interactively at export time (e.g.\ via a ``Save As'' dialog),
and therefore need not be named \texttt{observables.dat}. Likewise, GUI snapshots are saved to a user-chosen
\texttt{.json} path. Exported frame directories are selected by the user and then populated under a fixed
subdirectory name \\
\texttt{<basename(media\_path)>.export}.

\paragraph{Deterministic naming and lexical ordering.}
All per-frame artifacts should include a zero-padded frame index in the filename to preserve lexical ordering
and allow robust alignment with the numerical table. The current batch exporter writes exported images as
\texttt{\%05d.png} (e.g.\ \texttt{00000.png}, \texttt{00123.png}). Fractal-dimension diagnostic images and fit tables
are also frame-indexed, e.g.\ \texttt{fd\_fit\_d00042.dat} and \texttt{fd\_f00042\_...png}.

\paragraph{Portability across machines.}
To make runs portable across machines and re-runs, \texttt{PyPETANA} avoids embedding absolute paths inside output
tables. When paths are stored in JSON snapshots, the code records the \texttt{json\_path} relative to a
``media base directory'' (the parent folder of the video file, or the image-directory itself), and resolves it
when reloading (Sec.~\ref{sec:export_json}). This prevents path breakage when archiving and moving a run folder.

\subsection{Primary numerical output: frame-indexed observables table}
\label{sec:export_numeric}

The primary quantitative output of \texttt{PyPETANA} is a structured ASCII table containing frame-resolved
observables produced by the observables core (Sec.~\ref{subsec:observables}). Each row corresponds to one analyzed
frame index $i$ (time point), and columns store the geometric and fractal descriptors computed from the selected
morphology in that frame.

\paragraph{Default columns (current behavior).}
The default extraction writes a fixed set of columns:
\begin{itemize}
  \item \texttt{frame\_id}
  \item \texttt{area}, \texttt{perimeter}, \texttt{circularity}
  \item \texttt{fractal\_dimension}, \texttt{fractal\_dimension\_rms}
  \item \texttt{fractal\_dimension\_boundary}, \texttt{fractal\_dimension\_boundary\_rms}
\end{itemize}
If hole-aware observables are enabled in the analysis pipeline, the backend can additionally compute hole-adjusted
variants (Sec.~\ref{subsec:observables}). The on-disk schema above is intentionally minimal and portable for downstream
time-series analysis.

\paragraph{Lightweight provenance in the header.}
For reproducibility and bookkeeping, the header includes a machine-readable line of the form
\texttt{meta\_json=\{...\}}, containing key settings required to interpret the table. In the current implementation,
this metadata includes at least:
(i) \texttt{media\_basename},
(ii) the boundary-band thickness \texttt{contour\_thickness},
and (iii) the box-size series used for box counting \texttt{box\_sizes\_series}.
This lightweight provenance is designed to remain usable even when optional diagnostic artifacts are not exported.

\subsection{GUI snapshot JSON: \texttt{records.json}}
\label{sec:export_json}

\texttt{PyPETANA} exports a JSON file containing GUI-recorded parameter snapshots (``records'').
This file is intended for replaying and interpolating GUI settings across frames and therefore serves as the
canonical, machine-readable record of the preprocessing/segmentation configuration used during extraction.

\paragraph{File structure (current implementation).}
The exported JSON has the form:
\begin{lstlisting}
{
  "version": 1.1,
  "media_path": "...",
  "records": [ { ... }, { ... }, ... ],
  "json_path": "..."
}
\end{lstlisting}
Each element of \texttt{records} is a dictionary keyed by GUI labels (e.g.\ \texttt{(Select) Lower Threshold})
and includes a \texttt{Frame Index}. The record set can be sparse in time; during batch extraction, settings are
interpolated between recorded frames to produce a per-frame configuration.

\paragraph{Relative JSON path handling.}
To keep snapshots portable, \texttt{PyPETANA} stores \texttt{json\_path} relative to the media base directory
(the video’s parent directory or the image-directory itself). When loading, the stored relative path is resolved
against that media base directory. If relative paths cannot be constructed (e.g.\ Windows paths on different drives),
the code falls back to storing a normalized absolute path.

\paragraph{What this JSON guarantees (and what it does not).}
Given the same input media and the exported \texttt{records.json}, preprocessing/segmentation settings used during
extraction can be reproduced, including thresholds, blur/morphology, cropping, and tilt parameters, as well as
analysis toggles stored in the records. This JSON is not intended to be a full provenance report (e.g.\ OS,
Python version, CPU model, git commit, list of successfully processed frames). Instead, \texttt{PyPETANA} provides a
minimal provenance footprint in the numerical-table header via \texttt{meta\_json=\{...\}}
(Sec.~\ref{sec:export_numeric}).

\subsection{Diagnostic exports for validation}
\label{sec:export_diagnostics}

A central design goal of \texttt{PyPETANA} is that every numerical observable can be visually audited.
Accordingly, the export core can optionally write diagnostic artifacts that document segmentation quality,
contour selection, and the multiscale box-counting procedure.

\paragraph{Segmentation and contour overlays.}
When enabled, \texttt{PyPETANA} can export per-frame images that include:
(i) the cropped frame,
(ii) the thresholded image used for contour detection,
(iii) contour overlays (main contour and optional interior contours),
and (iv) optional masking, alpha background removal, and tilt visualization.
These overlays provide a direct link between pixel-level geometry and the corresponding rows in the observables table.

\paragraph{Fractal-dimension fit tables and occupancy patterns.}
When FD diagnostics are enabled, the export core may write per-frame tables
(e.g.\ \texttt{fd\_fit\_d\%05d.dat}) containing the log--log occupancy data used in the regressions
(Sec.~\ref{sec:fractal}), along with fit residuals and minimal metadata
(\texttt{meta\_json} including \texttt{contour\_thickness} and \texttt{box\_sizes\_series}).
Optionally, it can export images visualizing box occupancy at each scale and padding placement used for supersampling,
allowing users to verify the scaling window and assess discretization artifacts.

\subsection{How the GUI calls the backend}
\label{sec:gui_backend}

Each GUI update is dispatched through a single orchestrator routine
(\lstinline{update_region}), which:
(i) reads the current frame,
(ii) applies deterministic preprocessing/segmentation to produce contours and threshold images,
(iii) renders overlays for visual validation, and
(iv) optionally calls the observables routine \lstinline{extract_data} for the current frame.

Algorithm~\ref{alg:gui_orchestration} summarizes this event-driven flow.

\begin{algorithm}[H]
\caption{GUI orchestration (\lstinline{update_region}): event-driven dispatch to the \texttt{PyPETANA} backend}
\label{alg:gui_orchestration}
\KwIn{
GUI state (current frame index $i$; sliders/toggles);\;
session handles (\texttt{read\_frame}, \texttt{total\_frames});\;
\texttt{media\_path};\;
(export options and flags)
}
\KwOut{
Interactive previews (frames/masks/overlays);\;
(optional) per-frame observables displayed in the GUI
}

\tcp{1. Collect GUI parameters}
\texttt{params} $\leftarrow$ dict assembled from GUI controls (thresholds, blur, morphology, crop, tilt, \dots)\;

\tcp{2. Read current frame}
(\texttt{ok}, $I$) $\leftarrow$ \texttt{read\_frame(i)}\;
\If{\texttt{ok} is False}{
  \Return{}\;
}

\tcp{3. Preprocess + contour extraction}
$(I_{\mathrm{crop}},B,T,\mathcal{C},\mathcal{C}_{\star},x_{\mathrm{off}},y_{\mathrm{off}})$
$\leftarrow$ \texttt{preprocess\_frame}(I,\texttt{params})\;

\tcp{4. Visualization overlays (for GUI preview)}
\texttt{preview\_images} $\leftarrow$
\texttt{export\_frame}$(I,I_{\mathrm{crop}},T,\mathcal{C},\mathcal{C}_{\star},x_{\mathrm{off}},y_{\mathrm{off}},\texttt{params})$\;
Render \texttt{preview\_images} in the GUI\;

\tcp{5. Optional: compute observables for current frame}
\If{\texttt{compute\_observables}}{
  \texttt{row} $\leftarrow$
  \texttt{extract\_data}\big(
    B,$\mathcal{C}_{\star}$, \texttt{box\_sizes\_series}, i, \texttt{params}
  \big)\;
  Display selected fields of \texttt{row} in the GUI\;
}

\Return{}\;
\end{algorithm}

\subsection{Batch execution and multiprocessing}
\label{sec:batch_multiprocessing}

Batch extraction is executed by the backend routine \lstinline{extract_data_command}.
It (i) interpolates GUI \texttt{records} to a per-frame list of configurations,
(ii) optionally runs multiprocessing by chunking frames across workers, and
(iii) merges results into a single frame-indexed table written to the user-selected \texttt{.dat} output file.

Algorithm~\ref{alg:batch_multiprocessing} matches the current control flow\\
(\lstinline{extract_data_command} $\rightarrow$ \lstinline{process_interpolated_frames}).

\begin{algorithm}[H]
\caption{Batch extraction and multiprocessing (\lstinline{extract_data_command})}
\label{alg:batch_multiprocessing}
\KwIn{
\texttt{media\_path}; list of GUI \texttt{records}; global settings:
\texttt{include\_holes}, \texttt{contour\_thickness} ($\tau$), \texttt{box\_sizes\_series};\;
(optional) FD diagnostic toggles \texttt{extract\_fd\_fits}, \texttt{extract\_fd\_frames};\;
(optional) \texttt{num\_workers}
}
\KwOut{
User-selected \texttt{.dat} file (frame-indexed table);\;
User-selected \texttt{.json} file (GUI snapshot file);\;
(optional) FD diagnostics and/or FD visualization frames
}

\tcp{1. Interpolate GUI records}
Compute \texttt{interp\_configs} by interpolating sparse \texttt{records} across all frames\;

\tcp{2. Determine parallelism}
If \texttt{num\_workers} is not specified, set from available logical CPUs\;
Split \texttt{interp\_configs} into worker chunks\;

\tcp{3. Parallel map (optional)}
\If{\texttt{num\_workers} $> 1$}{
  Submit each chunk to \lstinline{process_interpolated_frames(media_path, chunk, ...)}\;
}
\Else{
  Run \lstinline{process_interpolated_frames} on the full list in the main process\;
}

\tcp{4. Reduce (merge results)}
Each worker returns a dict of column arrays (e.g.\ \texttt{area}, \texttt{perimeter}, \dots)\;
Concatenate returned arrays for each column key; sort rows by \texttt{frame\_id}\;

\tcp{5. Write table}
Write the merged structured array to the chosen \texttt{.dat} path with:
(i) a header listing column names and
(ii) a machine-readable \texttt{meta\_json=\{...\}} line containing at least
\texttt{media\_basename}, \texttt{contour\_thickness}, and \texttt{box\_sizes\_series}\;

\tcp{6. Optional FD diagnostics}
If enabled, write FD fit tables and/or FD occupancy visualizations under
\texttt{<media\_path>.extract/}\;

\Return\;
\end{algorithm}

\subsection{Reloading prior runs}
\label{sec:export_reload}

Previously exported JSON parameter files can be reloaded through the GUI to restore the analysis configuration used
in an earlier run. Reloading is designed to reproduce preprocessing, segmentation, and observables computation
without manual slider reconfiguration. To avoid accidental overwriting, we recommend writing re-run outputs into a
new output directory or using a run identifier subfolder (e.g.\ \texttt{outdir/run\_YYYYMMDD\_HHMMSS/}).

\paragraph{Reproducibility checklist.}
A complete \texttt{PyPETANA} analysis is reproducible given:
(i) the raw video or image sequence,
(ii) the exported JSON parameter snapshot (\texttt{records.json}),
(iii) the numerical output table (user-selected \texttt{.dat} file with \texttt{meta\_json=\{...\}} header),
and (iv) the archived code release (e.g.\ Zenodo) corresponding to the cited version.


\section{Benchmark suite}
\label{sec:benchmark}

The benchmark suite is designed to validate the numerical behavior of the observables
implemented in \texttt{PyPETANA} under controlled and reproducible conditions.
Because \texttt{PyPETANA} is formulated as a deterministic, geometry-first measurement
framework, the benchmarks focus on three complementary aspects:
(i) correctness of low-order geometric observables,
(ii) numerical stability of multiscale boundary-fractal estimation, and
(iii) reproducibility when applied to experimentally relevant evolving morphologies.

The purpose of the benchmark suite is not an exhaustive comparison against alternative
segmentation frameworks, but verification that the implemented measurements recover
expected morphological trends from fixed inputs and frozen parameter records.
Together, the benchmarks validate both compactness-sensitive geometric descriptors and
multiscale roughness observables associated with invasive morphology.

Unless otherwise noted, benchmark runs use a boundary thickness $\tau = 1$ pixel and
box-counting fits based on at least five distinct box sizes.
All segmentation settings are stored in frozen \texttt{records.json} files so that each
benchmark may be rerun deterministically from the same image and parameter set.

\subsection{Benchmark B1: Perimeter-sensitive geometric descriptors}
\label{subsec:benchmark_circularity}

\begin{figure}[H]
    \centering
    \includegraphics[width=0.95\linewidth]{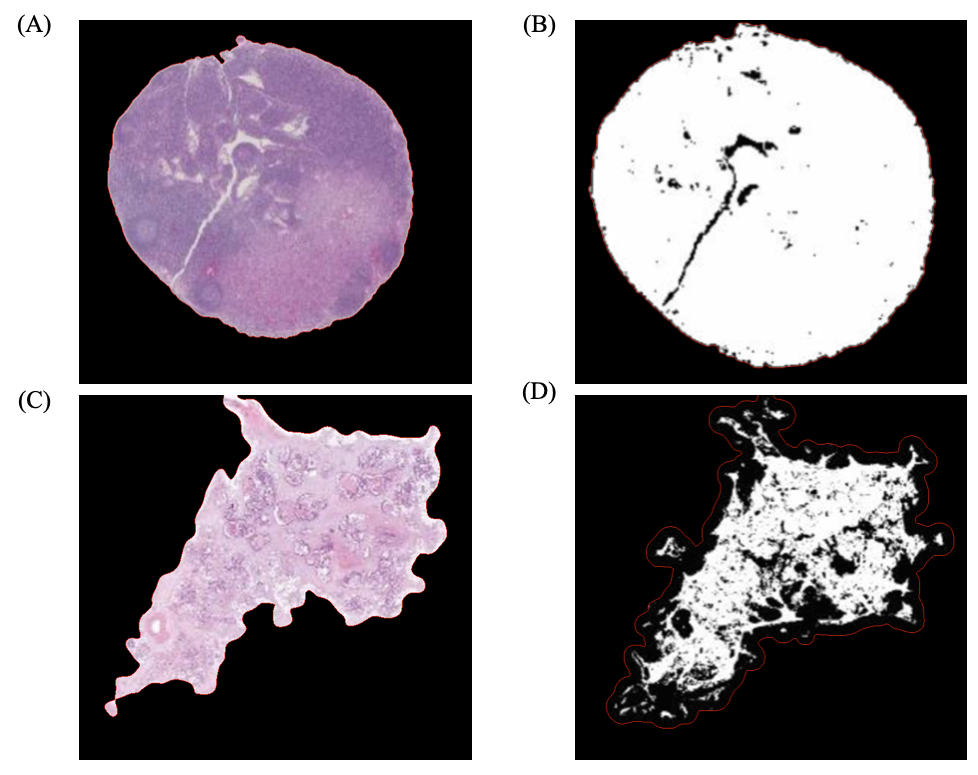}
    \caption{
Benchmark images from Ref.~\cite{brouwer2023} used to validate perimeter-sensitive geometric descriptors.
Panels (A) and (C) show representative segmented morphologies in tumor nodules\cite{brouwer2023}, while panels (B) and (D)
show the corresponding binary masks used for quantitative analysis together with the
extracted outer contour (red).
The top row corresponds to a compact morphology with an approximately circular boundary,
whereas the bottom row corresponds to a visibly invasive morphology with an extended and
irregular front.
}
    \label{fig:circ-benchmark}
\end{figure}

A fundamental requirement for any morphology-analysis framework is the ability to distinguish
compact from invasive growth forms using reproducible geometric observables derived directly
from contour geometry.
For compact morphologies, the perimeter remains close to that of a circle with the same
enclosed area, whereas invasive morphologies exhibit substantially larger perimeters relative
to area due to protrusions and boundary roughening.

This behavior may be quantified through the circularity~\cite{himberg2026geometrydynamicalmorphologygrowing}

\begin{equation}
C = \frac{4\pi A}{P^2},
\end{equation}

or equivalently through the complexity ratio~\cite{brouwer2023}

\begin{equation}
\chi = \frac{P}{2\sqrt{\pi A}} = C^{-1/2}.
\end{equation}

For an ideal circle, $\chi = 1$ and $C = 1$, while increasing boundary irregularity yields
$\chi > 1$ and $C < 1$.

Two representative morphologies related to tumor nodule were analyzed~\cite{brouwer2023}: a compact morphology
[Fig.~\ref{fig:circ-benchmark}(A,B)] and an invasive morphology
[Fig.~\ref{fig:circ-benchmark}(C,D)].
For each case, \texttt{PyPETANA} extracted the dominant outer contour from the binary mask
and computed the enclosed area $A$, perimeter $P$, and derived complexity ratio $\chi$.
Because all quantities are computed from the same contour representation, this benchmark
simultaneously validates contour extraction and downstream geometric measurement.

\begin{table}[h!]
\centering
\caption{Reference and measured complexity-ratio values for the benchmark morphologies.
Values close to unity indicate compact boundaries, whereas larger values correspond to
increasing boundary irregularity and invasive roughness.}
\begin{tabular}{lcc}
\hline
\textbf{Panel} & \textbf{Complexity Ratio$^{(\mathrm{Ref\cite{brouwer2023}})}$} & \textbf{Complexity Ratio$^{(\mathrm{PyPETANA})}$} \\
\hline
(A) & 1.06 & 1.116092 \\
(B) & 1.06 & 1.102705 \\
(C) & 1.68 & 1.819174 \\
(D) & 1.68 & 1.727983 \\
\hline
\end{tabular}
\label{tab:complexity_ratio}
\end{table}

The compact morphology yields complexity-ratio values close to unity
($\chi \approx 1.10$), consistent with a nearly circular outline.
By contrast, the invasive morphology yields substantially larger values
($\chi \approx 1.73$--$1.82$), reflecting the enhanced perimeter associated
with a roughened and protrusive interface.

Small quantitative deviations from the quoted reference values are expected due to
thresholding, rasterization, and contour discretization.
Nevertheless, the ordering of the morphologies is recovered unambiguously.
This benchmark therefore confirms that the deterministic contour-extraction procedure
implemented in \texttt{PyPETANA} reliably separates compact and invasive morphologies
through area- and perimeter-derived geometric observables.

\subsection{Benchmark B2: Boundary-fractal evolution of invasive tumor morphology}
\label{subsec:benchmark_mda}

Whereas Benchmark~B1 validates low-order geometric descriptors derived from contour geometry,
Benchmark~B2 probes the multiscale sensitivity of the supersampled box-counting implementation
through the extraction of the boundary fractal dimension $D_b$.

Tumor morphologies provide a particularly stringent benchmark for multiscale analysis because
invasive fronts simultaneously exhibit large-scale geometric deformation and fine-scale
boundary roughness across multiple spatial scales.
The present benchmark therefore evaluates whether \texttt{PyPETANA} captures the progressive
increase in invasive roughness associated with evolving tumor morphology.

The benchmark uses time-resolved MDA-MB-231 morphologies evaluated over four days of growth~\cite{10.1371/journal.pone.0109784}.
Boundary fractal dimensions were computed using the standard \texttt{PyPETANA}
boundary-band construction with thickness $\tau = 1$ pixel.
In addition to the standard analysis, a hole-aware variant was evaluated in which interior
voids retained in the contour hierarchy are incorporated explicitly into the boundary mask.
This allows direct assessment of the influence of interior topology on the fitted scaling behavior.

\begin{figure}[H]
    \centering
    \includegraphics[width=0.95\linewidth]{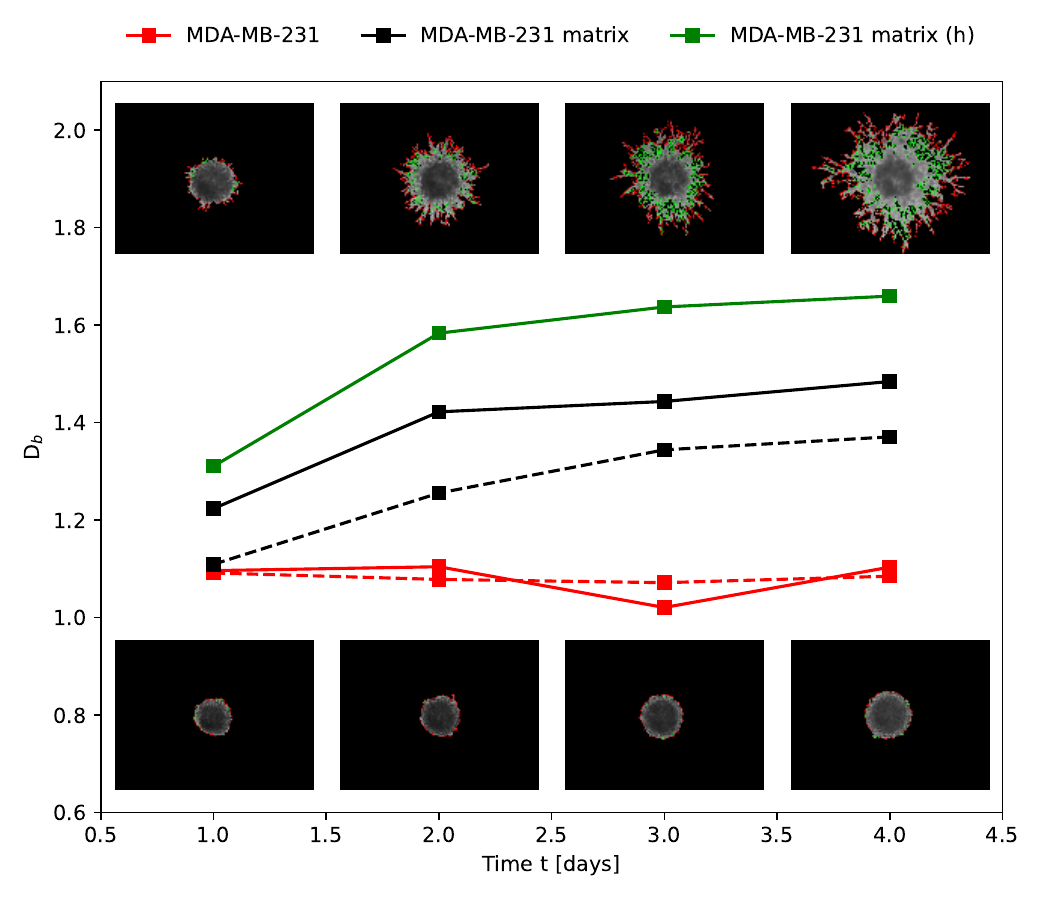}
    \caption{ Time evolution of the boundary fractal dimension $D_b$ for MDA-MB-231 morphologies
reported in Ref.~\cite{10.1371/journal.pone.0109784}.
The red dotted line corresponds to the directly measured morphology~\cite{10.1371/journal.pone.0109784},
while the black dotted line corresponds to the matrix-associated morphology~\cite{10.1371/journal.pone.0109784}.
The corresponding solid curves are computed using \texttt{PyPETANA}.
The green solid curve denotes the hole-aware matrix analysis performed within \texttt{PyPETANA}.
Insets show representative contours at successive times. The increase in $D_b$ for the matrix-associated morphologies indicates progressive growth of
boundary roughness, while the systematically larger hole-aware values reflect the additional
contribution of interior void structure to the effective multiscale boundary geometry.}

\label{fig:mda-benchmark}
\end{figure}

A clear separation is observed between comparatively smooth and invasive morphologies.
For the directly measured morphology, the fitted boundary fractal dimension remains near
$D_b \approx 1.05$--$1.10$ throughout the full time interval.
By contrast, the matrix-associated morphology exhibits a monotonic increase from
approximately $D_b \approx 1.22$ at day 1 to $D_b \approx 1.48$ at day 4.

When interior holes are retained explicitly, the measured dimension increases further,
rising from approximately $D_b \approx 1.31$ to $D_b \approx 1.66$ over the same interval.
These results demonstrate that the supersampled box-counting implementation resolves the
expected increase in multiscale boundary complexity associated with invasive growth.
Moreover, the hole-aware extension provides a controlled and interpretable shift in the
fitted scaling exponent arising from interior topology.

Taken together, Benchmarks~B1 and~B2 validate both the low-order and multiscale outputs of
\texttt{PyPETANA}.
Benchmark~B1 confirms that area- and perimeter-derived descriptors correctly distinguish
compact and invasive morphologies, while Benchmark~B2 demonstrates that the boundary-fractal
analysis captures the expected increase in invasive roughness and extends naturally to
hole-aware multiscale geometry.

\section{Limitations}
\label{sec:limits}

The analysis is primarily limited by image resolution, projection of the sample onto the two-dimensional plane, and sample isolation with respect to the substrate. Other factors include using a different version of OpenCV, currently constrained to version 4.12 for reproducibility, and fractal dimension settings such as box size series and contour width.

\texttt{PyPETANA} does not infer microscopic growth rules, agent dynamics,
or biochemical mechanisms.
It does not perform tracking of individual cells, lineage reconstruction,
or three-dimensional reconstruction.
These choices are intentional and reflect the geometry-first scope of the codebase.

\begin{itemize}
  \item \textbf{Illumination gradients:} if segmentation drifts across the field of view, enable luminance tilt correction.
  \item \textbf{Over/under-thresholding:} overly tight thresholds can fragment the colony; overly loose thresholds can merge background artifacts.
  \item \textbf{Empty or multiple components:} if no contour is detected or multiple large components exist, verify cropping and thresholds using exported overlays.
  \item \textbf{Crop boundary interactions:} if the morphology touches the crop boundary, measured perimeter/fractal estimates can be biased; expand the crop region.
  \item \textbf{FD fit quality:} always report the RMS residuals (\texttt{fractal\_dimension\_rms} and boundary RMS) as diagnostics of scaling consistency.
  \item \textbf{Grid alignment:} supersampled padding offsets reduce bias; if residuals remain high, revisit the box-size range and contour thickness.
\end{itemize}

\section{Conclusion}
\label{sec:conclusion}

\texttt{PyPETANA} provides a geometry-first, time-resolved framework for converting image data
into reproducible morphological observables.
By enforcing a strict separation between preprocessing/segmentation and the explicit mathematical
definition of observables, the framework produces frame-resolved measurements determined entirely
by the input media and frozen parameter records.
The tutorial presented here documents this mapping explicitly, from image frames to binary masks
and extracted contours, and ultimately to exported time series of $A(t)$, $P(t)$, $C(t)$, and
effective fractal dimensions for both filled morphologies and controlled boundary bands.
In addition, RMS residuals associated with box-counting fits are reported to quantify scaling quality
and assess the stability of the extracted exponents.

A central emphasis of \texttt{PyPETANA} is transparency, auditability, and reproducibility.
The exported numerical outputs (\texttt{observables.dat}) together with the GUI snapshot files
(\texttt{records.json}) provide a compact reproducibility footprint, while optional overlays and
fractal-diagnostic artifacts enable direct visual validation of segmentation, contour selection,
and multiscale counting assumptions on a frame-by-frame basis.

The framework is designed as a measurement protocol for evolving morphology rather than as an
inference engine for microscopic growth rules.
Its purpose is to provide consistent geometric descriptors that may be compared across experiments,
growth conditions, imaging modalities, and computational models.
Although the present work focuses primarily on biological morphologies, the benchmark suite further
demonstrates applicability to invasive tumor morphologies and multiscale boundary evolution in
time-resolved cancer-growth interfaces~\cite{10.1371/journal.pone.0109784,brouwer2023}.
More broadly, the underlying workflow is applicable to evolving non-equilibrium interfaces and
multiscale geometric patterns, including diffusion-limited aggregation, roughened growth fronts,
and interfacial instabilities encountered across non-equilibrium systems~\cite{witten1981diffusion,Barabasi1995}.

The physical interpretation of these observables in the context of bacterial colony growth is
discussed separately in Ref.~\cite{himberg2026geometrydynamicalmorphologygrowing}.
The present tutorial instead focuses on the computational workflow itself, with the goal of making
the full analysis pipeline transparent, reproducible, and extensible for future applications.

\section*{Code availability}
\texttt{PyPETANA} is released under an open-source license GPLv3 and archived on Zenodo.
The exact version corresponding to this tutorial is cited in Ref.~\cite{sanghita_pypetana_2024}.

\section*{Acknowledgements}
We acknowledge support from Brandeis University and computational resources
from the Vermont Advanced Computing Center.

\section*{Funding information}
The authors received no specific funding.


\bibliographystyle{unsrturl}
\bibliography{refs_bh}

\end{document}